\documentclass[namedreferences]{solarphysics}
\usepackage[hyperref,optionalrh]{spr-sola-addons} 
\usepackage{graphicx}                    
\usepackage{url}                         
\usepackage{amssymb}                     
\usepackage{lscape}                      
\usepackage{color,soul}

\makeatletter
\newcommand*{\rom}[1]{\expandafter\@slowromancap\romannumeral #1@}
\makeatother

\begin{document}

\begin{article}

\begin{opening}

\title{{\bf He {\sc i} vector magnetic field maps of a sunspot and its superpenumbral fine-structure}}

%
\author{T.A.~\surname{Schad}$^{1,4}$\sep
       M.J.~\surname{Penn}$^{2}$\sep
       H.~\surname{Lin}$^{1}$\sep
       A.~\surname{Tritschler}$^{3}$}

%
\runningauthor{Schad et al.}
\runningtitle{The Chromospheric Vector Magnetic Field Structure of a Sunspot}

%
\institute{$^{1}$ Institute for Astronomy, University of Hawai`i, Pukalani, HI 96768
                    email: \url{schad@ifa.hawaii.edu}, \url{lin@ifa.hawaii.edu}  \\ 
           $^{2}$ National Solar Observatory, 950 N. Cherry Avenue, Tucson, AZ 85719 email: \url{mpenn@nso.edu}\\
           $^{3}$ National Solar Observatory, 3665 Discovery Drive, Boulder, CO 80303 email: \url{ali@nso.edu}\\
           $^{4}$ Previously at the Department of Planetary Sciences of the University of Arizona, with joint affiliation with the National Solar Observatory. }

\begin{abstract} 
Advanced inversions of high-resolution spectropolarimetric observations of the He {\sc i} triplet at 1083 nm are used to generate unique maps of the chromospheric magnetic field vector across a sunspot and its superpenumbral canopy. The observations were acquired by the {\it Facility Infrared Spectropolarimeter} (FIRS) at the {\it Dunn Solar Telescope} (DST) on 29 January 2012.  Multiple atmospheric models are employed in the inversions, as superpenumbral Stokes profiles are dominated by atomic-level polarization while sunspot profiles are Zeeman-dominated but also exhibit signatures perhaps induced by symmetry breaking effects of the radiation field incident on the chromospheric material.  We derive the equilibrium magnetic structure of a sunspot in the chromosphere, and further show that the superpenumbral magnetic field does not appear finely structured, unlike the observed intensity structure. This suggests fibrils are not concentrations of magnetic flux but rather distinguished by individualized thermalization. We also directly compare our inverted values with a current-free extrapolation of the chromospheric field.  With improved measurements in the future, the average shear angle between the inferred magnetic field and the potential field may offer a means to quantify the non-potentiality of the chromospheric magnetic field to study the onset of explosive solar phenomena.
\end{abstract}

\end{opening}


\section{Introduction}\label{sec:intro} 

Mapping the upper atmospheric solar magnetic field is essential to understanding the quasi-continual nonthermal heating of the chromosphere/corona. Around sunspots and other field concentrations, the ability of the magnetic field to either suppress or enhance energy transport from the photosphere into the upper atmosphere is being more widely recognized, especially now with the transition region diagnostics of the {\it Interface Region Imaging Spectrograph} (IRIS) (see, {\it e.g.}, \inlinecite{depontieu2014}).  This is despite a sincere lack of magnetic field diagnostics in the upper atmosphere.  Most observational studies of the influence of the upper atmosphere magnetic field on nonthermal heating have relied on extrapolations of the photospheric field. \inlinecite{mcintosh_judge2001}, in this manner, located `magnetic shadows' thought to be closed field regions of the chromosphere that suppress upward propagating MHD waves.  Complementary, \inlinecite{depontieu2005} found photospheric p-mode oscillations may be leaked or channeled into the corona by a reduced effective gravity along inclined magnetic fields, perhaps driving dynamic fine-scaled fibrils and jets around sunspots \citep{hansteen2006}. 

Spectropolarimetric measurements of these fine-scaled chromospheric features are only starting to probe their vector magnetic structure \citep{delaCruz2011,schad2013}.  Difficulties in the measurement and interpretation of linearly polarized spectral lines formed partially in the chromosphere led early studies to utilize only \textit{longitudinal} chromospheric magnetograms \citep{choudhary2001,jin2013}, which give a mixed view of the magnetic environment hosting fibrils.  For example, \inlinecite{giovanelli1982} argued fibrils denote local variations in the gas excitation of a more uniform magnetic canopy, while \inlinecite{zhang1994} suggested fibrils were regions of concentrated magnetic flux. 

Mapping the upper atmospheric magnetic field is also essential to understand large-scale evolution of active regions and the implusive release of mass and energy in solar flares and coronal mass ejections (CMEs).  Studies of the magnetic structure of active regions have again mostly been limited to the photosphere (see the review by \inlinecite{solanki2003review}), whereas chromospheric magnetic conditions within active regions have been probed with vector magnetometry limitedly within the strong field regions directly above sunspots and their penumbra \citep{socas_navarro2005a,socas_navarro2005b,orozco_suarez2005, delacruz2013} or in solar prominences/filaments \citep{casini2003,kuckein2012a,orozco2014}.  Yet, mapping the full chromospheric magnetic field above active regions is potentially very useful.  We suspect due to the relaxed morphology of the chromosphere's intensity structure that the stresses of the magnetic field dominate over those of the plasma ({\it i.e.}, a low plasma $\beta$).  Some evidence is given by \inlinecite{metcalf1995}.   Extrapolations of the coronal magnetic field should improve with field measurements in such a force-free environment, since current methods with and without preprocessing of the photospheric field to enforce the force-free condition are inconsistent \citep{derosa2009}.

Here we demonstrate recent advancements in the measurement and interpretation of the He {\sc i} triplet at 1083 nm, formed in the upper chromosphere.  We derive unique maps of the chromospheric magnetic field vector extending from within a sunspot's umbra out through its superpenumbra, which demarcates a sunspot's outer chromospheric boundary \citep{bray1974}.  Our work is an extension of that of \inlinecite{schad2013} (hereafter, Paper {\rom 1}), which determined the full field vector along the spines of individually resolved superpenumbral fibrils.  Measurable polarized signals both in linear and circular polarization were found throughout the sunspot and its superpenumbra, ranging from a strong-field Zeeman-dominated regime directly within the sunspot to atomic-level polarization dominated signatures within the superpenumbral canopy and beyond. Inversions of the Stokes spectra to infer the magnetic field thus require the quantum-level spectropolarimetric modeling of \inlinecite{asensio_ramos2008}.  Application of this model distinguishes the maps derived and analyzed here (Section~\ref{sec:results}) from earlier studies ({\it cf.}, \inlinecite{lagg2004}, \inlinecite{orozco_suarez2005}, and \inlinecite{bloomfield2007}). 

\begin{figure}
\centerline{\includegraphics[width=0.45\textwidth]{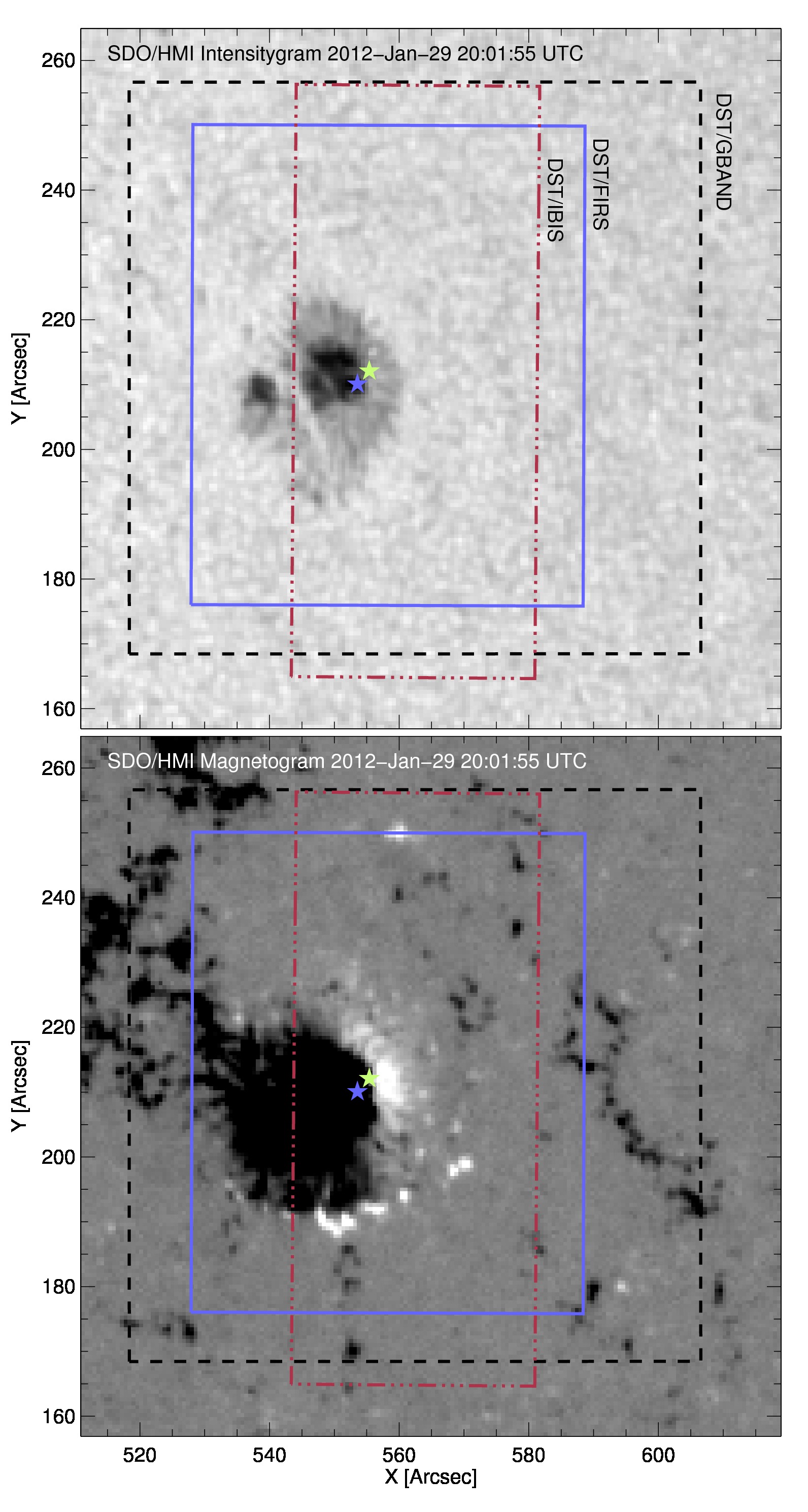}}
\caption{Regions of NOAA active region 11408 targeted by the instruments at the {\it Dunn Solar Telescope} on 29 January 2012 overplotted on and referenced to the SDO/HMI intensitygram (top) and longitudinal magnetogram (bottom) acquired at 20:01:56 UTC.  The axes give helioprojective coordinates referenced as seconds of arc away from disk-center.  He {\sc i} Stokes profiles gleaned from the sunspot (blue and yellow stars) are shown in Figure~\ref{fig:inv_models_umb}.}
\label{fig:fov_obs}
\end{figure}


\section{Observations}\label{sec:obs_reduce}

We analyze very high sensitivity measurements of the He {\sc i} triplet polarized spectra at 1083 nm obtained with the {\it Facility Infrared Spectropolarimeter} \citep[FIRS:][]{jaeggli2012} during multi-instrument observations at the DST.  Paper \rom{1} described in detail the observational setup used for these observations, acquired on 29 January 2012.  For completion, we give a brief description here.  The best determination of the chromospheric vector fields comes from a single deep-integration FIRS slit-scan across NOAA active region (AR) 11408.  The map consists of 200 individual steps of a $77''$ project length slit oriented parallel with the solar central meridian (as viewed from Earth) and scanned from solar East to West between 19:16 and 20:50 UTC.  Slightly East of the center of the FIRS maps is a single unipolar sunspot, located at N8W35 ($\mu = \cos{\Theta} = 0.8$; see Figure~\ref{fig:fov_obs}).  After full post-facto reduction, the spatial sampling is $0.3'' \times 0.3''$.  Long integrations of 7.5 seconds at each slit position resulted in mean noise levels in the polarized Q,U, and V spectra of  $4.0\times10^{-4}$, $3.7\times10^{-4}$, and $3.0\times10^{-4}$, in units of continuum intensity ($I_{C}$), respectively.  Filtered 2D principal component analysis helped remove residual interference fringes \citep{casini2012}.  These spectra have been corrected for instrumental polarization (estimated error $\lesssim 0.076\%$ $I_{C}$), and the wavelength scale has been calibrated in an absolute sense by accounting for Sun-Earth orbital motions, solar rotation, and the solar gravity redshift.  Therefore, the reported Doppler velocities correspond to an absolute reference frame fixed to the solar surface (estimated error of $\pm 250$ m sec$^{-1}$.)

Using standard image processing techniques, we coalign the FIRS observations with an intensitygram and longitudinal magnetogram acquired during the FIRS scan by the {\it Helioseismic and Magnetic Imager} \citep[HMI:][]{scherrer2012} on board NASA's {\it Solar Dynamics Observatory} (SDO).  Figure~\ref{fig:fov_obs}, which shows the field-of-view of all operating DST instruments, is the reference for the continuum structure underlying the FIRS maps discussed in Section~\ref{sec:results}.  A $5.12 ' \times 5.12'$ portion of the full disk HMI magnetogram is used below (see Section~\ref{sec:field_shear}) to extrapolate the magnetic field into the upper atmosphere under the current-free approximation.


\section{Analysis of the He {\sc i} Triplet Polarization}\label{sec:he_inv}

\begin{figure}
\centerline{\includegraphics[width=0.99\textwidth]{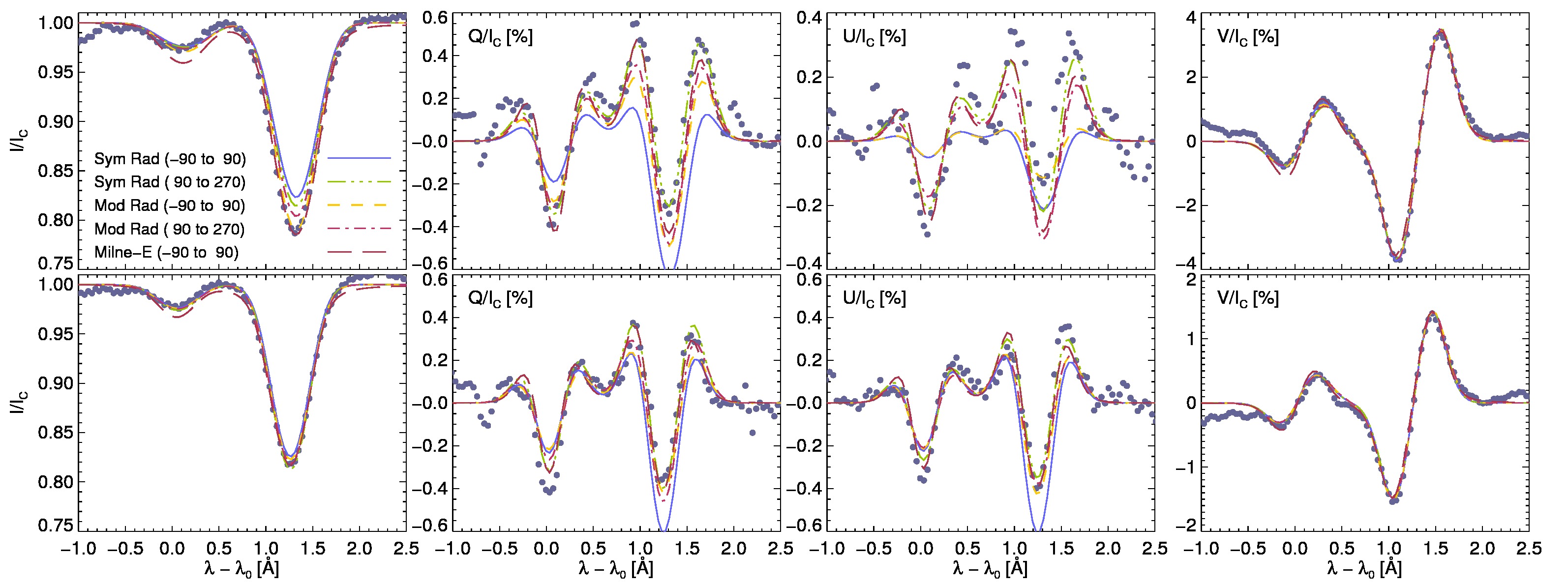}}
\caption{Example umbral (top row) and penumbral (bottom row) He {\sc i} Stokes spectra extracted from the spatial locations identified by the blue star and yellow star in Figure~\ref{fig:fov_obs}, respectively, are given by the data points.  {\sc Hazel} model fits resulting from constant property slab models with nominal (solid and triple-dot-dashed lines) and modified (short-dashed and dot-dashed lines) prescribed incident radiation fields are shown alongside a Milne-Eddington model (long dashed line) without atomic level polarization.  Goodness-of-fit parameters are given in Table~\ref{tbl:umb_models}.}
\label{fig:inv_models_umb}
\end{figure}

A number of mechanisms generate and/or modify polarization in the spectrum of the orthohelium triplet ({\it i.e.}, He {\sc i} 1083 nm) \citep{trujillo_bueno_2007, asensio_ramos2008}. Paper \rom{1} identified largely Zeeman dominated Stokes profiles directly above the sunspot umbra and penumbra, while the superpenumbral canopy, wherein 39 resolved fibrils were identified, presented polarized spectra clearly dominated by atomic-level polarization.  By atomic-level polarization, we refer to polarization induced by population imbalances of and coherences between the magnetic sublevels of each atomic term in the orthohelium system.  Inferring the magnetic field parameters within this regime requires inversions whose forward engine models the emerging Stokes spectra using the full quantum framework.  This means solving first the statistical equilibrium equations governing the atomic density matrix for an ensemble of helium atoms subject to an anisotropic radiation field \citep{landi_2004}.  

For the superpenumbra, our inversions employ the Hanle and Zeeman Light diagnostic tool known as {\sc Hazel}, which is a forward modeling and inversion code developed by \inlinecite{asensio_ramos2008} based on the multiterm calculations of \inlinecite{landi_2004}.  The radiative transfer equation of the forward model consists of a constant-property slab model including magneto-optical terms and stimulated emission.  The slab's characteristics, and secondarily the modeled and fit Stokes profiles, are quantified by a thermal Doppler broadening, $v_{th}$, a macroscopic line-of-sight (LOS) velocity, $v_{mac}$, an optical thickness, $\Delta\tau$, damping parameter, $a$, magnetic field intensity, $B$, magnetic field inclination angle, $\theta_{B}$, magnetic field azimuthal angle, $\chi_{B}$, and a height, $h$.  A newer parallelized version of the code, known as {\sc P-Hazel}, is well suited for generating maps.  Parallelization is essential; each single-point inversion ({\it i.e.}, each pixel) takes from a minute to a few minutes to complete using a single modern processor and the relatively-efficiently optimization scheme implemented by {\sc P-Hazel}.  For an alternative, promising approach, see the PCA-based optimization scheme used by \inlinecite{casini2003}.

\subsection{The Influence of Symmetry Breaking Effects in the Umbra}

There is, however, a limitation to using {\sc Hazel} for interpreting the polarized spectra formed in the strong field regions of the sunspot umbra and inner penumbra, one that for this particular sunspot led us to adopt a Milne-Eddington model in this region.  The limitation stems from its prescription of the pumping radiation field.  The degree of atomic-level polarization in the atomic ensemble of helium atoms in the solar atmosphere is controlled in part by the anisotropic properties of the incident radiation field.  In its current implementation, {\sc P-Hazel} specifies the pumping radiation field incident on a slab of helium atoms at a given height with a `classical' unpolarized atmosphere, meaning it represents a feature-less cylindrically-symmetric angle-dependent photospheric continuum flux.  This radiation field is fully quantified by two nonzero components of the irreducible spherical radiation field tensor, $J_{Q}^{K}(\nu)$, defined as:
\begin{equation}
J_{Q}^{K}(\nu) = \oint \frac{d\Omega}{4\pi}\sum_{i=0}^{3}{T}_{Q}^{K}(i,\vec{\Omega})S_{i}(\nu,\vec{\Omega}),\label{eqn:rad_field_tensor}
\end{equation}
where $S_{i}$ is the radiation field vector along a ray vector and ${T}_{Q}^{K}$ are irreducial spherical tensors that depend on the reference frame and ray direction (see Section 5.11 of \inlinecite{landi_2004} for further details).  Only the $K = 0, 2$ and $Q = 0$ components are necessary to describe the simple classical atmosphere, which greatly reduces the number of statistical equilibrium equations that must be solved in the forward model to calculate the atomic-level polarization.  However, in a solar active region, the presence of features such as plage, sunspots, or pores, break the cylindrical symmetry.  As a consequence, to define the radiation field tensor at each frequency $\nu$ associated with a transition in the atomic model requires multi-spectral knowledge of the localized radiation field.  Beyond this complication, incorporating the additional multipole moments of the radiation tensor greatly increases the complexity of the model leading to even longer forward-calculation time.

Initially, we attempted to apply the {\sc Hazel} model without consideration of the influence of the additional multiple moments of the radiation field. As a first exercise, we employed {\sc Hazel} for the inversion of the polarized spectra observed within the sunspot umbra and penumbra using completely open bounds for the magnetic field direction ({\it i.e.}, all azimuths from $0$ to $360^{\circ}$ were allowed).  Surprisingly, the majority of the inverted values of the magnetic field azimuthal angles in the superpenumbra represented a diverging sunspot magnetic field instead of the expected converging field for a negative polarity sunspot (see Figure~\ref{fig:fov_obs}).  This was not a consequence of the optimization algorithm, but rather the diverging field solution more adequately fit the Stokes profiles than the diverging field solution.  

\begin{table}
\caption{{\bf Comparison of Umbral Inversion Models.} Magnetic field vector strength, inclination and azimuth are specified in the LOS coordinate frame, where inclination is measured relative to the line of sight and the azimuth is referenced as degrees counterclockwise from the solar West direction. $\chi^{2}_{I,Q,U,V}$ refer to the reduced chi-squared goodness-of-fit values for Stokes I,Q,U, and V, respectively.}
\label{tbl:umb_models}
\scriptsize
\begin{tabular}{cccccccc}
\hline
Model & $ B [G]$ & $\Theta_{B} [^\circ]$ & $\Phi_{B} [^\circ] $ & $\chi^{2}_{I}$ & $\chi^{2}_{Q}$ & $\chi^{2}_{U}$ & $\chi^{2}_{V} $ \\
\hline
\multicolumn{5}{l}{Outer Umbral Spectrum - Blue Star} \\
 1 &  1463.6 & 139.8 &                  187.1  & 2.85 & 40.2 & 23.0 & 12.1 \\
 2 &  1711.7 & 127.0 & \phantom{0}15.5   & 1.67 &  2.0 &  5.7 & 15.6 \\
 3 &  1395.5 & 131.2 &                    184.7  & 0.45 & 13.9 & 18.8 & 14.8 \\
 4 &  1592.1 & 128.6 &  \phantom{0}14.3   & 0.93 &  9.1 & 12.7 & 13.7 \\
 5 &  1322.9 & 134.1 &                     184.6  & 0.37 & 14.4 & 19.1 & 13.8 \\
 6 &  1584.6 & 129.0 &  \phantom{0}14.3   & 0.94 & 10.5 & 13.6 & 13.5 \\
 7 &  1363.2 & 127.1 &  \phantom{0}15.0  & 0.68 &  5.4 &  9.0 & 44.3 \\
\hline
\multicolumn{5}{l}{Inner Penumbral Spectrum - Yellow Star} \\
 1 &  1272.8 & 109.7 &                  201.8   & 0.41 & 11.1 & 11.5 &  5.5 \\
 2 &  1322.2 & 107.0 & \phantom{0}19.9   & 0.23 &  2.1 &  1.9 &  5.5 \\
 3 &  1269.3 & 109.9 &                   202.3   & 0.21 &  5.4 &  4.9 &  4.7 \\
 4 &  1306.4 & 108.7 &  \phantom{0}19.1  & 0.16 &  4.3 &  4.4 &  4.6 \\
 5 &  1275.8 & 109.9 &                   202.3  & 0.21 &  4.9 &  4.3 &  4.7 \\
 6 &  1292.6 & 108.3 & \phantom{0}25.8   & 0.18 &  4.8 &  4.2 &  4.8 \\
 7 &  1174.2 & 108.9 & \phantom{0}21.5  & 0.50 &  2.3 &  2.5 &  7.0 \\
\hline
\end{tabular}
\end{table}

To better illustrate the challenges posed by the umbral spectra, we perform inversions of two different representative Stokes spectra with seven different forward models.  Figure~\ref{fig:fov_obs} indicates the locations of the observed spectra, which are displayed alongside selected model fits in Figure~\ref{fig:inv_models_umb}.  Models 1 thru 6 employ the constant-property slab solution of the radiative transfer equation as in \inlinecite{asensio_ramos2008}, while model 7 uses a Milne-Eddington atmospheric model.  For the Milne-Eddington model, we employ the {\sc Helix$^{+}$} inversion code with its {\sc Pikaia} genetic optimization \citep{lagg2004,lagg2007}.  Model 7 then includes the Zeeman and Paschen-Back effects on the polarized He {\sc i} spectra, but no effects of atomic-level polarization.  Model 1 and 2 use {\sc Hazel} in its nominal mode where the radiation field is cylindrically-symmetric.  The only difference is that the field solutions are restricted to different ranges of azimuthal angles in a local reference frame orthogonal to the solar surface (No. 1: $-90^{\circ} < \chi_{B} < 90^{\circ}$ and No. 2: $90^{\circ} < \chi_{B} <270^{\circ}$).  Models 3 and 4 are similarly defined by these portions of the magnetic field directional space, but for these models the $J_{0}^{0}$ and $J_{0}^{2}$ components of the radiation field tensor are modified in accordance with the direct calculation given in Appendix.  Models 5 and 6 neglect atomic-level polarization entirely.  The resulting magnetic field vectors and reduced $\chi^{2}$ fit values are reported in Table~\ref{tbl:umb_models}.

The outer umbral spectrum (i.e. the blue star) exhibits modification of its linear polarization due to atomic-level polarization, particular in Stokes U, wherein the $\pi$ lobe of the red triplet component is less strongly polarized then the $\sigma$ lobes.  This effect was similarly witnessed by \inlinecite{lagg2004} and explained by \inlinecite{trujillo_bueno_2007} as signs of atomic level polarization.  Alternatively, the inner penumbral spectra shows a definitive lack of atomic-level polarization influence.  Stokes Q and U for both Stokes spectra are nearly symmetric, suggesting the $K = 1$ components of the radiation field are negligible.  That is, the pumping radiation field does not carry net circular polarization.  Also in both cases, the Milne-Eddington model (No. 7) better fits the linear polarization than the converging field solutions with atomic level polarization ({\it i.e.}, Nos. 1,3,5).  The best-fit to Q and U in both cases is the diverging field solution for Model 2, as discussed above.  In the outer umbral spectrum, the fit for Model 2 is considerably better than most fits despite this field being physically difficult to conceive.  We argue, rather, that the non-negligible influence of the radiation symmetry breaking may anomalously lead to the preference of Model 2 although it is unphysical.  We find in Appendix that the $Q \ne 0$ components of the radiation field tensor (especially $Q = \pm  1$) within the sunspot are the same order of magnitude as the $Q = 2$ component in the nominal quiet classical atmosphere.  An alternative explanation could be that a constant-property slab model, as used by {\sc Hazel}, may inadequately describe He {\sc i} formation in the umbra. 

Despite great variations in the goodness of fit for the various radiative models (Table~\ref{tbl:umb_models}), the inferred field vector (with $180^{\circ}$ ambiguity) remains constant to within $\approx 20\%$, which has already been pointed out by \inlinecite{trujillo_bueno_2007}.  This remains true throughout the sunspot umbra due to the strength of the Zeeman effect.  Outside of this region, the atomic-level signatures begin to dominate but the level of symmetry breaking also greatly decreases. 

To quantify the degree of symmetry breaking in the radiation field pumping the He {\sc i} material, an estimate for the not-well-constrained height of the fibrils is necessary.  Contribution heights of He {\sc i} formation range between $\approx$ 1500 to 2200 km according to non-local thermodynamic equilibrium (non-LTE) calculations \citep{fontenla1993}.  Meanwhile, Paper \rom{1} found a high level of correspondence between He {\sc i}, H$\alpha$, and Ca {\sc ii} fibrils.  The numerical models of \inlinecite{leenaarts2012} argue that H$\alpha$ fibrils form at heights ranging from 1500 to 2750 km.  In accordance with these studies, we select a nominal height of 1750 km for the height of the He {\sc i} material in and around the observed sunspot.

In the Appendix we directly calculate the level of symmetry breaking within the radiation field tensor incident on the He {\sc i} material.  However, simple geometrical considerations supply an estimate for the lateral extent of the sunspot's influence.  The half angle of the heliographic extent for the light cone impinging on the raised material is given by:
\begin{eqnarray}
\varphi  & = & \frac{\pi}{2} - \gamma \\  \nonumber
	 & = & \frac{\pi}{2} - \arcsin\frac{R_{\odot}}{R_{\odot} + h} 
\end{eqnarray}
where $\gamma$ is the angle between the solar vertical at the point of the material and the ray tangent to the solar surface.  Thus, at a height of 1750 km, $\varphi$ is $4.06^{\circ}$, meaning that the material located more than $\varphi$ heliographic degrees away from the symmetry breaking source is unaffected.

\subsection{Multi-Model inversions}

Inside the sunspot, the inclusion of atomic polarization without extension of {\sc P-Hazel} to include symmetry breaking effects may give misleading results.  Consequently, we select the nominal mode of {\sc P-Hazel} for the inversion of spectra only outside of the sunspot.  Following previous authors (e.g. \inlinecite{bloomfield2007},\inlinecite{orozco_suarez2005}), we use the Milne-Eddington model as implemented in {\sc Helix$^{+}$} for the inversion of the He {\sc i} triplet spectra where the mean intensity of the pumping radiation, quantified by $J_{0}^{0}$, is reduced by more than 10$\%$ (see black region in Figure~\ref{fig:inv_ambig}). 

For both models, we use a single magnetized atmospheric component to fit the Stokes spectra.  The height of the slab used by {\sc P-Hazel} is assumed to be 1750 km above the solar surface, as in Paper \rom{1}. As shown by \inlinecite{asensio_ramos2008}, errors in the assumed height manifest primarily as errors in the inferred magnetic field inclinations.  However, due to the observational noise in these measurements, small differences in the assumed height do not affect the results presented here.  A number of pixels exhibit more complex profiles, with multiple, blended velocity components.  We identified the pixels whose He {\sc i} Stokes profiles are better represented by multiple atmosphere components using the Milne-Eddington model ({\sc Helix$^{+}$}) to fit one and two component model fits to the Stokes I and V spectra only.  Application of the Bayesian information criterion for model selection (see \inlinecite{ asensio_ramos_2012}) finds that less than one percent of the profiles in the observed region needs to be fit with multi-component atmospheres.  We do not perform further fits of the these pixels, and eliminate them from the remaining analysis.  We further ignore all pixels that do not contain a measurable polarization signal (at least $2\sigma$ above the noise) in at least one of the Stokes states.  Profiles without a measurable Stokes V are still included if the linear polarization is measured.  In the superpenumbra, this leads to areas in which the magnetic field intensity is not well constrained, as discussed below.  The liberal $2\sigma$ cutoff is used instead of a more conventional $3\sigma$ cutoff in order to investigate the weak profiles nearest the outer edge of the superpenumbral fibrils.  While the inversions of these noisy spectra produce spatially coherent maps of the magnetic field vector, the noisy distribution in the values prohibit any conclusive statements regarding the nature of the other fibril endpoints.

\subsection{Saturated Hanle Effect Ambiguities}

\begin{figure}
\centerline{\includegraphics[width=0.99\textwidth]{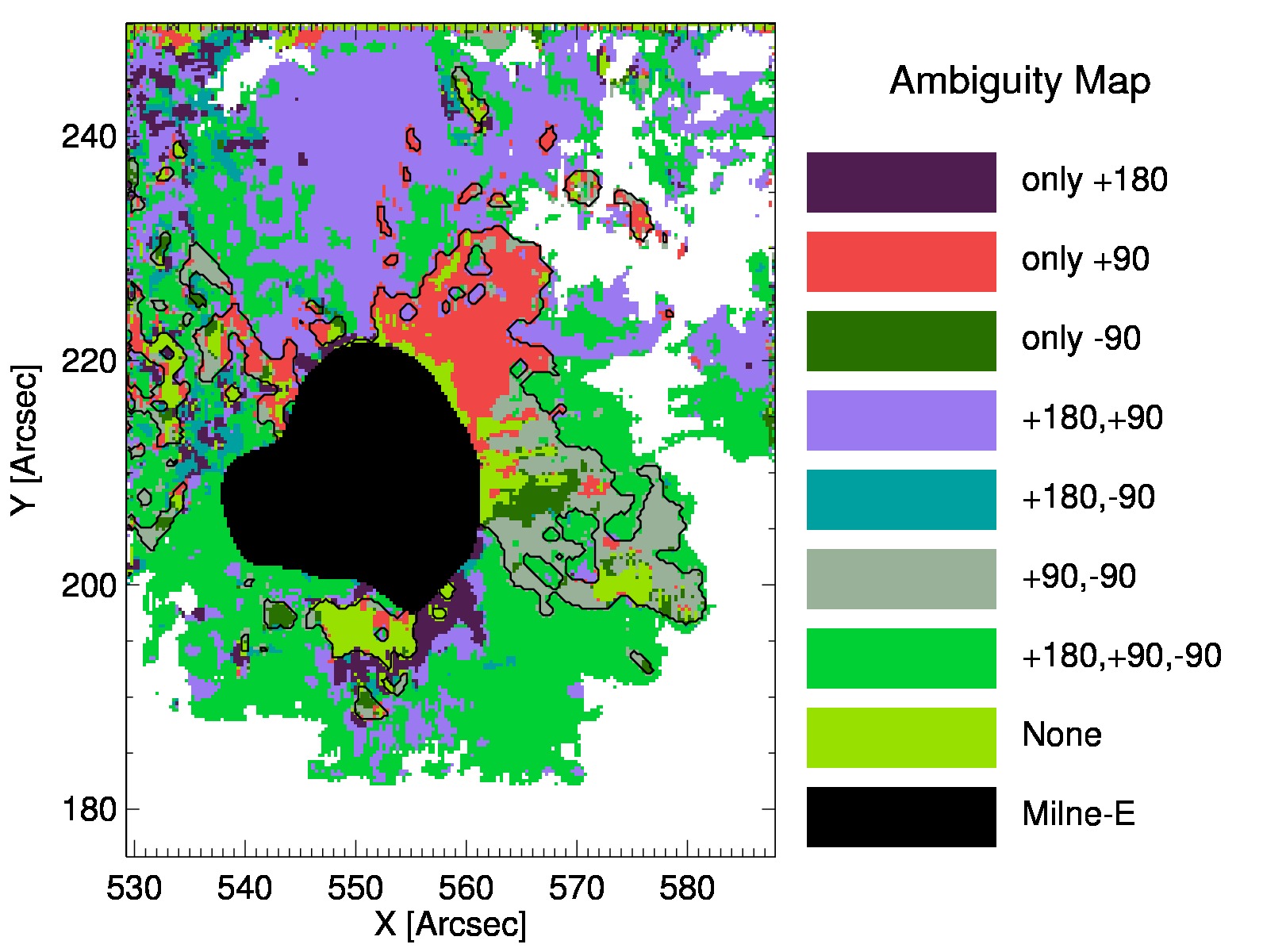}}
\caption{Regions of inversion ambiguities across the observed map.  The ambiguities result from the combined effects of the saturated Hanle effect ambiguities (i.e. the $180^{\circ}$ and $\pm 45^{\circ}$ Van Vleck LOS azimuth ambiguities) and the $180^{\circ}$ azimuthal ambiguity of the transverse Zeeman effect in the Milne-Eddington inversion subregion.  The contour outlines the region where no $180^{\circ}$ ambiguity exists. }
\label{fig:inv_ambig}
\end{figure}

To create coherent maps via inversions of our observations, we must also handle the presence of real physical ambiguities.  Just as the transverse Zeeman effect contains an inherent $180^{\circ}$ azimuthal ambiguity (in the line-of-sight reference frame), use of atomic-level diagnostics is also affected by ambiguities.  Paper \rom{1} described the possible ambiguities introduced within a classical unpolarized atmosphere for the regime in which the coherences of the atomic ensemble have been destroyed in the magnetic field reference frame.  This regime, known as the \textit{saturated Hanle regime}, contains up to eight possible ambiguities for the determination of the magnetic field direction via the linear polarized Stokes vectors (Q and U).  With a measurable Stokes V, this number reduces to four, two introduced by the $180^{\circ}$ ambiguity, and up to two introduced by the Van Vleck effect\footnote{See Section 5.8 of \inlinecite{landi_2004} for a discussion of the Van Vleck angle.}.  The Van Vleck effect ambiguities are not resolved with Stokes V since the magnetic field strength can be scaled to fit Stokes V in the saturated Hanle regime; Stokes Q and U are unaffected by the field strength up until to point that the transverse Zeeman effect is significant \citep{merenda2006}.  

Yet, the primary Hanle ambiguities do not apply for all field directions (see \inlinecite{merenda2006}) and observational geometries (see Paper \rom{1}).  This is a consequence of the preferential axis of the Zeeman effect signal being the line of sight, while that of the atomic-level polarization is the radiation symmetry axis.  For example, Stokes V often resolves the $180^{\circ}$ Hanle ambiguities for oblique scattering geometries on the solar disk since the other solutions would imply a vastly different orientation with respect to the symmetry axis.  For the saturated Hanle regime in a classical atmosphere, the four primary ambiguous solutions (with measurable Stokes V) are not immediately identified by the location of the best-fit solution.  According to \inlinecite{landi_2004}, these solutions correspond to linear polarization being oriented either parallel or perpendicular to the projection of the magnetic field vector along the LOS, yet the inclination and magnetic field strength of these solutions remain to be determined.  

We add an additional step to the inversion process to find these possible solutions by keeping fixed the thermodynamic parameters of the best-fit solution and fixing the LOS azimuth ($\Phi_{B}$) of the magnetic field vector in accordance with the saturated Hanle effect ambiguities.  That is, we transform the best-fit {\sc P-Hazel} solution into LOS coordinates with  $\Phi_{B,best}$ being the best-fit solution for the azimuthal field direction.  The four primary ambiguous solutions are then: $\Phi_{B,best}$,    $\Phi_{B,best} + \pi$, $\Phi_{B,best} + \pi/2 $, and $\Phi_{B,best} - \pi/2$.  Modifying {\sc P-Hazel} such that $\Phi_{B}$ is held constant, one iteration of the DIRECT algorithm followed by a fine-tuning iteration of the Levenburg-Marquart algorithm is used to fit two variables: magnetic field strength B, and the inclination in the LOS geometry $\Theta_{B}$.  Once the best-fits are found, they must be verified as ambiguities.  We only consider the additional solutions to be ambiguities if the reduced chi-squared values for Stokes Q, U, and V each remain minimally changed ($\Delta\chi^{2} < 1$).  A map of the regions affected by the various combinations of ambiguities is given in Figure~\ref{fig:inv_ambig}.  Note that no $180^{\circ}$ ambiguity exists for the region northwest of the sunspot.  The large Stokes V and the orientation of the fibril fields in this region resolve this ambiguity for that region (see Paper \rom{1}). 

\subsection{Ambiguity Resolution}

``Disambiguation" refers to the process of \textit{choosing} the ambiguous solution that best represents the solar magnetic field subject to assumptions regarding the physical nature of these fields.  At photospheric heights, numerous methods have been applied to real and artificial vector magnetograms with various levels of success \citep{metcalf2006}.  Similarly, maps of the magnetic field in the chromosphere must be ``disambiguated'' by choosing the most physically relevant solution.  We note that for our observed map, {\sc P-Hazel} does not produce a speckled mess of solutions despite the fact that we do not restrict the range of possible magnetic field directions during the inversion process.  In most areas, the solution is smooth across the region before disambiguation.  This is likely a consequence of the shape of the goodness-of-fit space coupled with the manner in which DIRECT searches for solutions.  Furthermore, the ambiguous solutions that we locate are rarely strict ambiguities.  For the oblique scattering angle, strict ambiguities with identical goodness of fit values may not exist except due to the role of observation noise.  

\begin{landscape}
\begin{figure}
\centerline{\includegraphics[height=0.7\textheight]{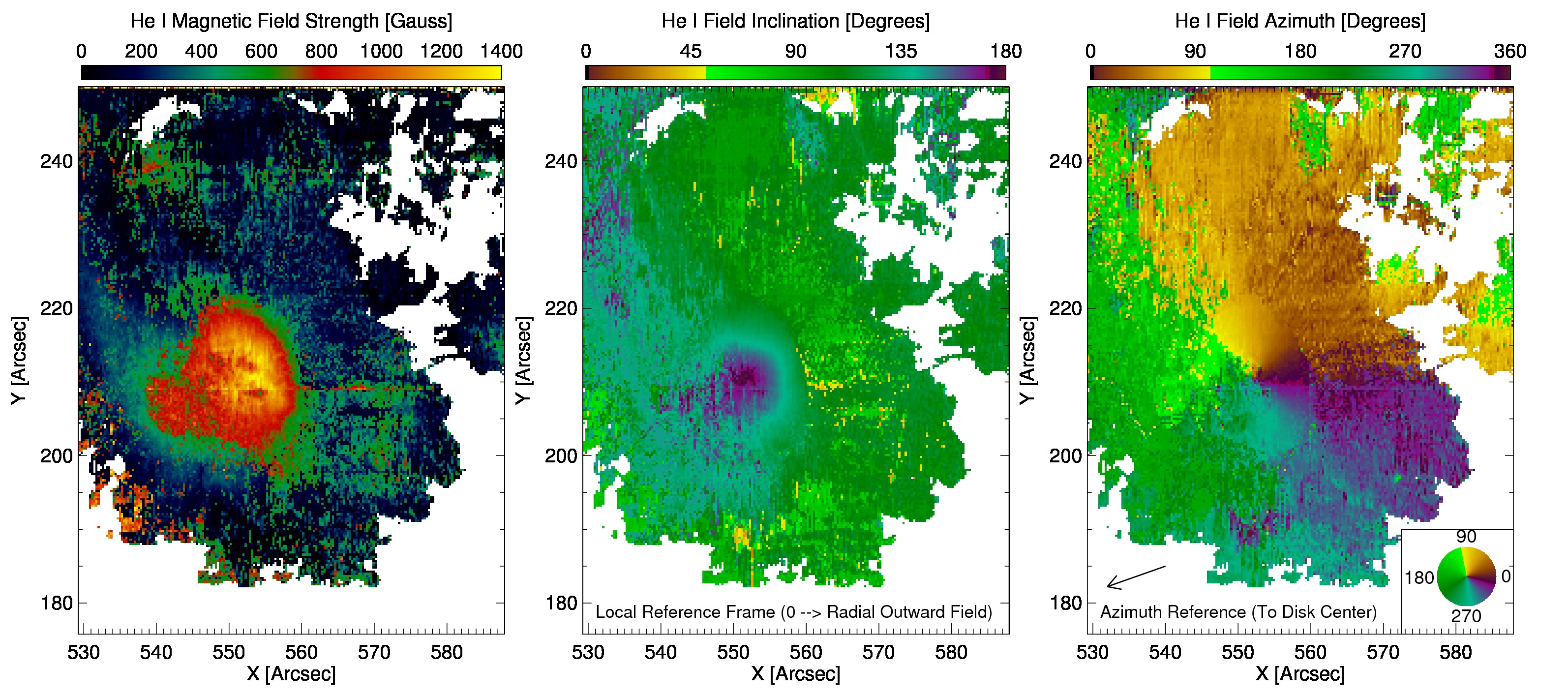}}
\caption{Maps of the magnetic field vector across an active region sunspot and its superpenumbral canopy inferred via analysis of the He {\sc i} triplet polarized spectra.  Ambiguities have been resolved across the region so too best represent a smooth continuous converging field solution.  The field inclination and azimuth are given in the local reference frame.  Solar North is up.  Solar West is to the right. }
\label{fig:field_solutions}
\end{figure}
\end{landscape}

To resolve the He {\sc i} ambiguities, we developed an interactive routine similar to the AZAM utilty developed for the {\it  Advanced Stokes Polarimeter} instrument \citep{lites1995}.  The two primary assumptions made are that the field converges for the negative polarity sunspot and that neighboring pixels give a continuous magnetic field solution.  Maps of the ``disambiguated" magnetic field are given in Figure~\ref{fig:field_solutions}, exhibiting a smooth converging field solution with a continuous superpenumbral canopy exhibiting the expected azimuthal change in the field direction around the sunspot.  The Milne-Eddington solutions inside the sunspot and the {\sc P-Hazel} solutions throughout the canopy exhibit remarkable spatial continuity at their boundary. 

\section{Results} \label{sec:results}

\subsection{The Chromospheric Magnetic Structure of a Sunspot} \label{sec:chromo_mag}

In Figure~\ref{fig:field_solutions}, we show the ambiguity-resolved maps of the chromospheric magnetic field vector across the sunspot and the superpenumbra.  The magnetic field structure of the superpenumbra exhibits remarkable continuity on the West and North sides of the sunspot.  The third panel of Figure~\ref{fig:field_solutions} gives the azimuthal angle of the magnetic field in the local frame where the angle is measured relative to the direction of disk center.  Thus, the field is converging into the sunspot and wraps around the magnetic field with a nearly radial orientation consistent with the field of fibrils studied by Paper \rom{1}.  The strength of the magnetic field drops from near 1500 Gauss in the sunspot umbra to between 100 and 300 Gauss within the superpenumbral canopy.  The superpenumbral canopy is confirmed to be a canopy by the inclination maps, showing values near $90^{\circ}$, which corresponds to a vector parallel to the solar surface.  

Note that the magnetic field strengths on the northwest side of the sunspot in Figure~\ref{fig:field_solutions} are less noisy than on the southwest side.  The 100 to 300 Gauss field strengths place the formation of the He {\sc i} polarized spectra within the saturated Hanle effect regime, but the transverse Zeeman effect does not give an appreciable signal for these field strengths.  As the linear polarization in the saturated regime does not depend on the magnetic field strength, the Stokes V profile determines the field strength.  Thus, the field strength is better determined in areas where the structure presents a larger longitudinal component to the observer, explaining why the northwest side field strengths are more coherent.  The fibrilar structure on the southwest side of the sunspot also exhibits more complex structuring in the {\it Interferometric Bidimensional Spectropolarimeter} (IBIS) observations of the DST in Ca \rom{2} 854.2 nm (not shown), which may also lead to greater level of noise apparent in the magnetic field strengths measured in that region.  

\begin{figure}
\centerline{\includegraphics[width=0.99\textwidth]{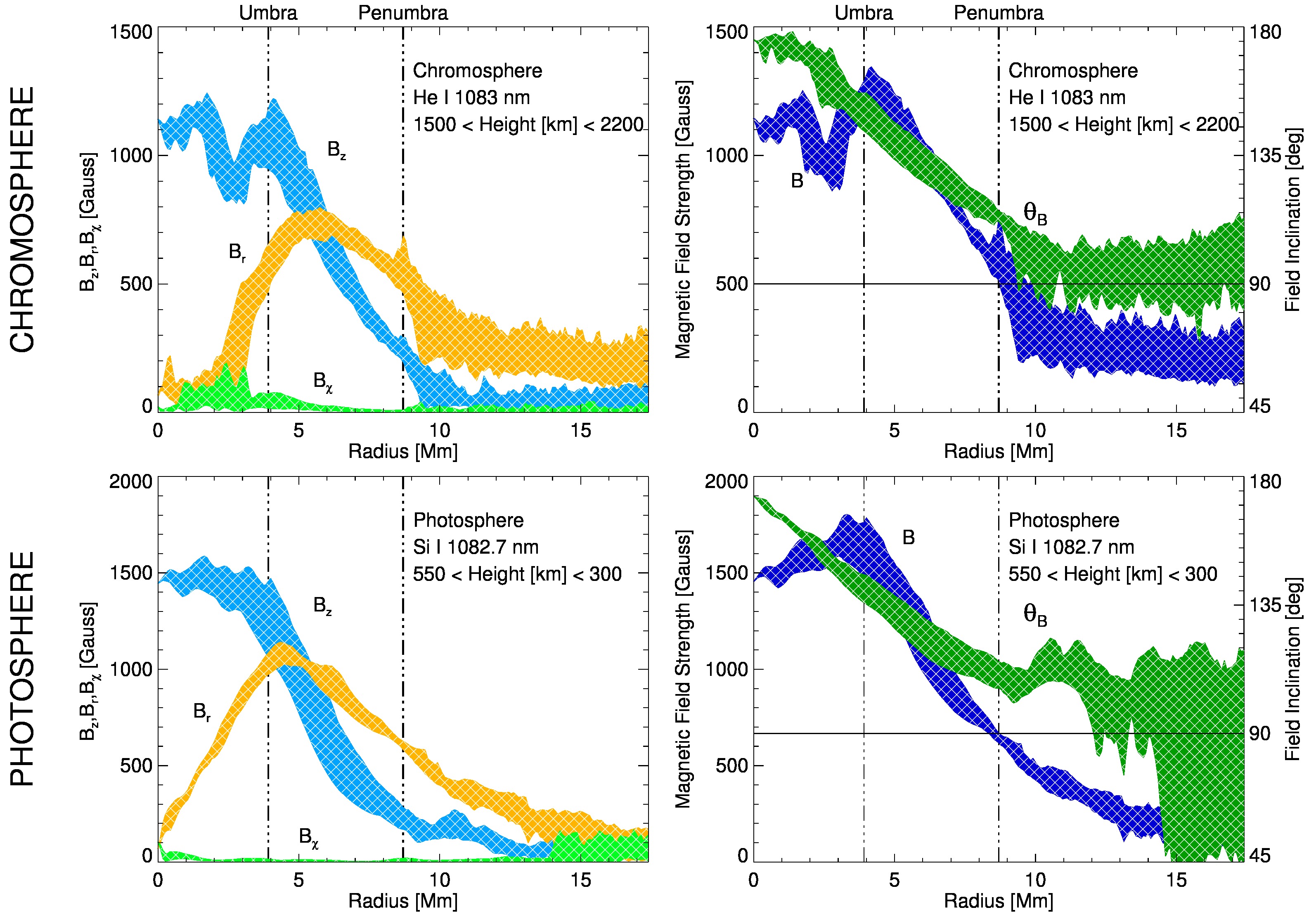}}
\caption{The average magnetic structure of the sunspot in the chromosphere and photosphere, resulting from an analysis of the He {\sc i} triplet at 1083 nm and the Si {\sc i} line at 1082.7 nm, respectively. $B_{z}$,$B_{r}$,and $B_{\chi}$ (left plots) give the strength of the magnetic field components in a cylindrical reference frame centered on the sunspot, while $B$ and $\theta_{B}$ (right plots) show the magnetic field's average total strength and inclination angle (relative to the solar vertical).}
\label{fig:hzhel_brbz}
\end{figure}

The overall structure of the sunspot in the chromosphere is assessed in Figure~\ref{fig:hzhel_brbz}, which shows averages of the field parameters in the azimuthal direction circumscribing the sunspot.  Using a cylindrical geometry centered on the nearly circular sunspot (see Paper \rom{1}), we represent the average structure of the sunspot as a function of sunspot's radius, just as in previous studies of the average photospheric structure of sunspots \citep{beckers1969, keppens1996}. Due to the error introduced in the magnetic field strength by the weak Stokes V signal discussed above, we study the average structure only of the northwest portion of the sunspot in Figure~\ref{fig:hzhel_brbz}.  We also compare the average chromospheric magnetic structure with the photospheric magnetic structure inferred from a one-component Milne-Eddington analysis of the Si {\sc i} spectral line at 1082.7 nm measured by FIRS (see Paper \rom{1}).  

The equilibrium magnetic structure of the sunspot in the chromosphere and photosphere shown in Figure~\ref{fig:hzhel_brbz} displays significant similarities.  In particular the magnetic field strength decreases as a function of radius while the inclination of the field decreases ({\it i.e.}, the field becomes more horizontal).  The presence of a light bright in the umbra (see Figure~\ref{fig:fov_obs}) leads to the relatively flat average behavior of the photospheric magnetic field in the umbra.  The chromospheric magnetic field, however, shows a dip in the magnetic field strength within the umbra.  In this portion of the umbra, the He {\sc i} Stokes Q and U spectra do not show clear symmetric profiles, and are characterized by an increase in noise.  Likely the high-frequency oscillations in the dark umbra leads to the disruption of the profiles over the long integration time of the FIRS observations.  Still, the approximately 500 Gauss reduction of the field strength in the umbra between the Si {\sc i} and He {\sc i} values is consistent with an approximately $0.5$ Gauss km$^{-1}$ vertical magnetic field gradient, assuming a 1000 km difference in the formation heights of the two lines in the umbra \citep{centeno2006}.  It is also clear that the ``magnetic radius'' of the sunspot in both the photosphere and chromosphere extend well beyond the outer edge of penumbra viewed in the continuum intensity. 

For this observed sunspot, we do not find any significant twist of the magnetic field for either the photospheric or chromospheric observations since the average $B_{\chi}$ component remains negligible (see Figure~\ref{fig:hzhel_brbz}).  Thus the torsional forces do not largely change in an average sense as a function of height.  By `twist' and `torsion', we refer to the same definitions used by \inlinecite{socas_navarro2005a}, where twist characterizes the non-radial behavior of the field in the cylindrical geometry centered on the sunspot, and torsion refers to the vertical gradient of the azimuthal direction of the field.  On a finer scale, clear changes in the azimuthal angle between the photosphere and chromosphere are seen, suggestive of complicated torsional forces at work within the sunspot itself.  However, since the observations cannot establish the formation height for the He {\sc i} triplet, projection effects due to the oblique observation angle of these observations can lead to perceived torsional forces that may or may not be real.  

Unfortunately, while disk center observations reduce the impact of projection effects, the horizontal nature of the superpenumbral canopy presents a challenge for the He {\sc i} triplet to infer magnetic field strengths since the linearly polarized signal is very weakly influenced by the magnetic field intensity.  Our observations more firmly constrain the field strengths in the superpenumbral canopy due to the oblique observation geometry.  The superpenumbral canopy fields observed here express a radial gradient smaller than the complementary canopy fields of the upper photosphere measured by Si {\sc i} (right panels of Figure~\ref{fig:hzhel_brbz}).  Outside of the penumbral boundary, the photospheric field continues to decrease rapidly while the chromospheric field becomes nearly constant. 

\begin{figure}
\centerline{\includegraphics[width=0.99\textwidth]{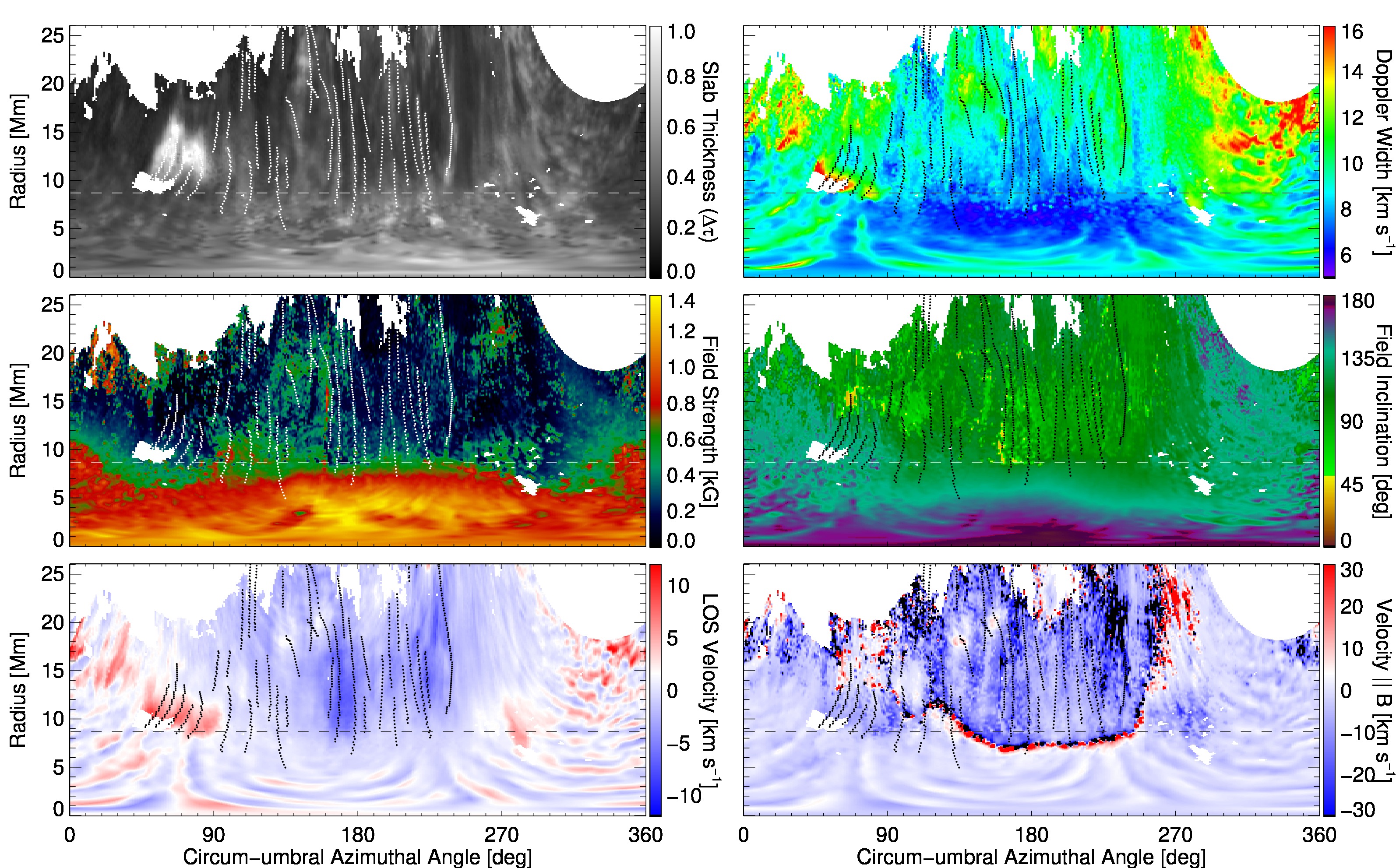}}
\caption{Magneto-thermal structuring of the superpenumbral region.  The observed FIRS maps is projected onto the cylindrical coordinate system centered on the sunspot, with the dashed line representing the outer penumbral edge.  White regions denote areas not well fit by the inversions or not measured by the FIRS scan.  Dotted lines trace the selected fibrils modeled in Paper \rom{1}.  The panels refer to the quantify indicated by the color bars, with the lower right panel giving the flow velocity magnitude assuming the flow follows the magnetic field lines.}
\label{fig:fibril_deproj}
\end{figure}

\subsection{Fibril and Inter-Fibril Magnetic Structure of the Superpenumbral Canopy}

In absence of vector field diagnostics for the regions surrounding sunspots, earlier studies attempted to deduce the magnetic structure of the superpenumbra using longitudinal magnetograms in chromospheric spectral lines.  \inlinecite{giovanelli1982} argued using Ca {\sc ii} 854.2 nm and H$\alpha$ magnetograms that the superpenumbral region consisted of a diffuse, but uniform, canopy of magnetic fields extending from the sunspot with a base height of less than 700 km.  Noting that the magnetograms did not contain fine-structuring consistent with the fibril structures seen in intensitygrams, the authors concluded that the superpenumbral fibrils do not delineate changes in the field structure, but rather changes in the gas structure and/or excitation.  Later work by \inlinecite{zhang1994} disputed this claim, illustrating fine scale structures in H$\beta$ magnetograms corresponding to dark filamentary structures extending from a sunspot.  This article claimed superpenumbral fibrils were regions of concentrated magnetic flux.  However, interpreting chromospheric filter-based longitudinal magnetograms requires not only consideration for projection effects but also for fine scale structuring produced by spatial velocity gradients.  With the full vector measurements presented here, we more adequately address the magnetized structure of both the fibrils and the inner fibril plasma.  

Figure~\ref{fig:fibril_deproj} displays the inferred magneto-thermal structure of the observed active region from the core of the sunspot umbra out to three times its penumbral radius.  The images have been projected onto coordinate axes in the cylindrical reference frame centered on the sunspot.  The x-axis gives the circumumbral azimuthal angle referenced to the direction of disk center.  That is, the region of the image between $90^{\circ}$ and $270^{\circ}$ corresponds to the half of the sunspot furthest from disk center.  The inward (blue-shifted) Doppler velocities measured in this portion of the sunspot (see lower left panel in Figure~\ref{fig:fibril_deproj}) are consistent with the inverse Evershed effect \citep{evershed1909}.  Dotted lines illustrate the paths of the 39 fibrils manually traced in Paper \rom{1} by inspecting the full spectral data cube.  Additional lineated structures not traced in Paper \rom{1} are apparent in the left top panel showing the inverted He {\sc i} slab thickness ($\Delta\tau$).  It is important to note that the He {\sc i} triplet exhibits measurable absorption across the entire observed scan, not just within the fibrils.  The fibrils only appear in contrast as regions with greater absorption, most likely due to greater mass densities (see Paper \rom{1}).

Overall, the magnetic field parameters of Figure~\ref{fig:fibril_deproj} (middle panels) support a more uniform magnetic architecture for the superpenumbral region compared to its fine intensity structure; albeit, the level of noise in the observations does lead to considerable scatter in the returned magnetic field parameters.  Still, our measurements indicate that the intensity structuring in the superpenumbra most likely does not result from the concentration of the magnetic field.  It is though surprising to find that the fibrils are not easily identified in the Doppler width panel (top right).  While there is a gradient towards larger natural line widths with increased distance from the sunspot, the Doppler temperature of the fibrils seems to be only slightly larger than the inner-fibril material.  

\begin{figure}
\centerline{\includegraphics[width=0.99\textwidth]{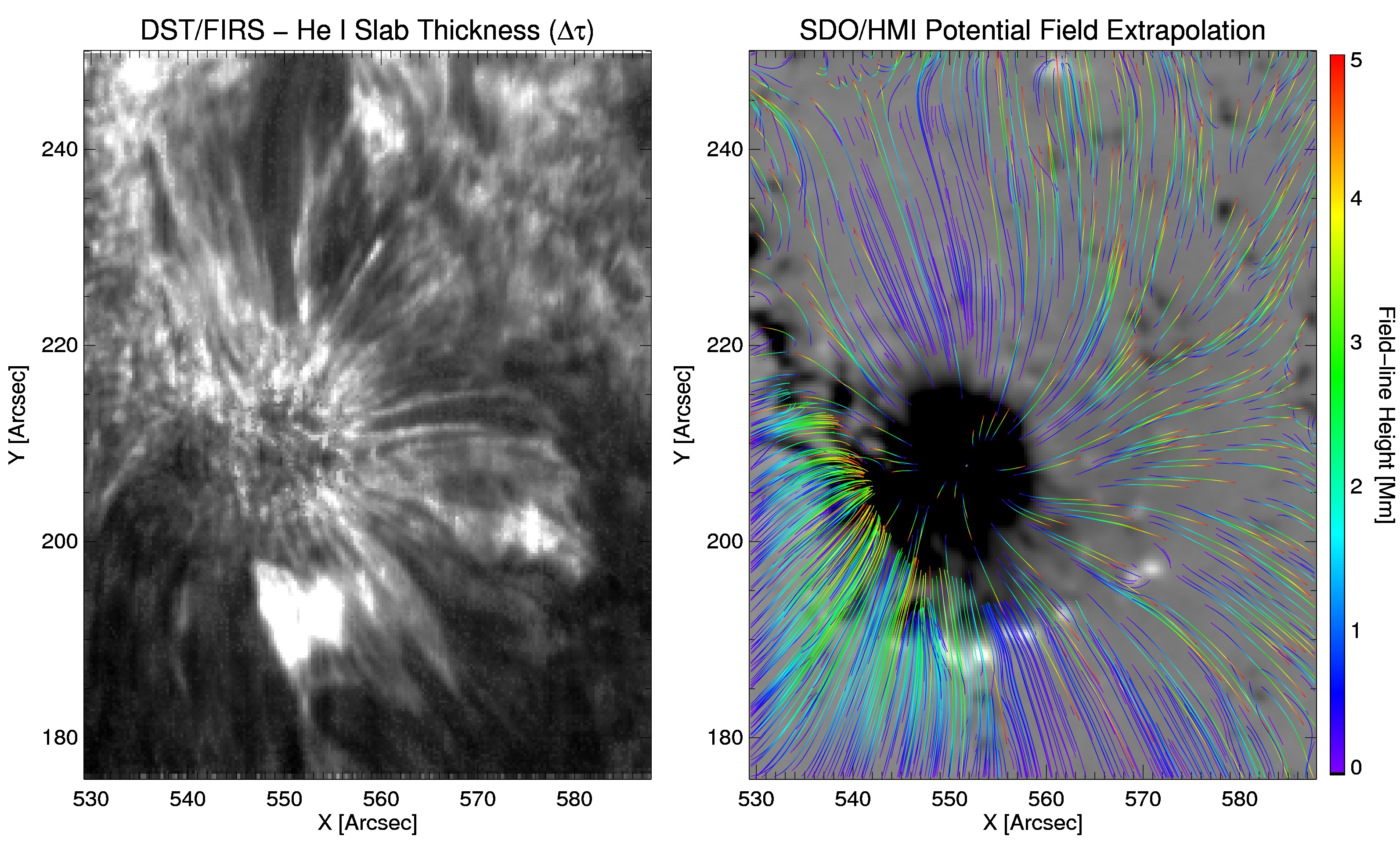}}
\caption{A comparison of the chromosphere fine intensity structuring (left) and the lines of force in a current free potential field extrapolation of the photospheric magnetic field (right).  The left panel gives the inferred slab thickness from {\sc P-Hazel} inversions of the observed He {\sc i} Stokes profiles.  Only field lines extending up to 5 Mm above the solar surface are shown on the right. Solar North is up.}
\label{fig:field_extrap}
\end{figure}

\subsection{Towards Quantifying Non-Potential Energy in the Chromosphere}\label{sec:field_shear}

With a full determination of the magnetic field vector in the chromosphere via the He {\sc i} triplet, a measure of the non-potentiality of the region may be obtained by calculating the shear angle between the inferred magnetic field \textit{vector} and the extrapolated current-free field.  By using the full measured vector, this method expands that of \inlinecite{jing2011}, who performed a similar exercise by calculating the mean shear angle between the azimuthal component of the extrapolated field and the projected azimuthal direction of fine-scaled fibrils observed in pseudo-monochromatic intensitygrams.  However, two issues complicate the application of this method.  First, we have made an assumption in the inversions regarding the height of the He {\sc i} material, placing it at 1750 km.  Likely, the He {\sc i} observations represent not a uniform level of the atmosphere, but rather a corrugated sheet.  Until we establish a better constraint for the height, the shear angle may give anomalous values.  Second, despite the advanced instrumental capabilities demonstrated here, our measurements still contain a high degree of observational noise relative to the measured signals.  Determining the true errors in the inverted quantities remains an area of research.  Thus, here we only compare the potential field extrapolation to our inverted quantities rather than quantify the magnetic shear angle.

\begin{figure}
\centerline{\includegraphics[width=0.45\textwidth]{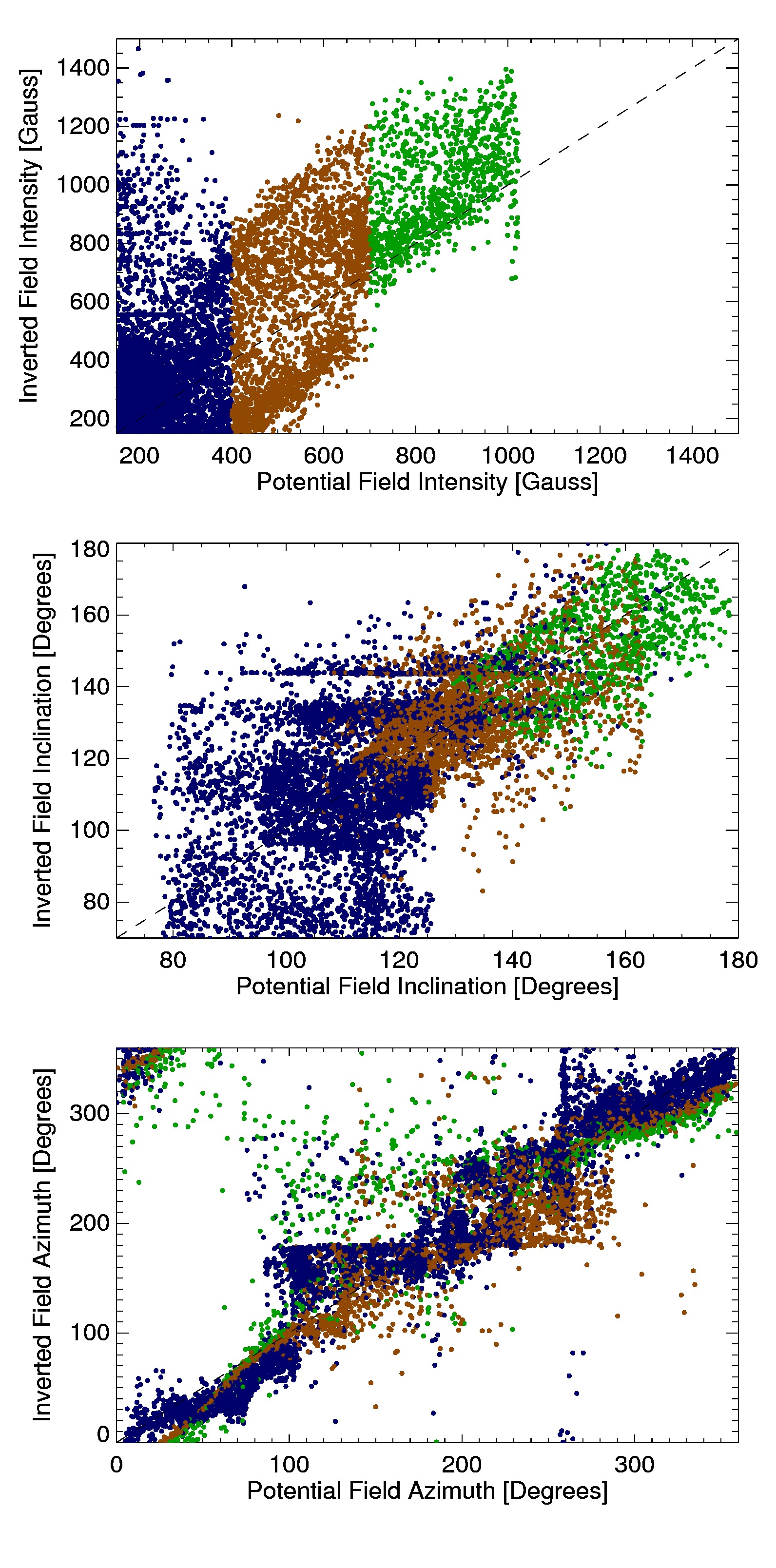}}
\caption{Comparison of the chromospheric magnetic field intensity (top), inclination (middle), and azimuth (bottom) for the inverted map and a current-free extrapolation of SDO/HMI magnetograms.  Inclination and azimuths are given in the local reference frame of the sunspot.  Data points are colored according to three ranges in the potential magnetic field intensity (see top panel).} 
\label{fig:pot_field_shear}
\end{figure}

We perform a current-free extrapolation of the photospheric magnetic field using a $5.12' \times 5.12'$ subregion of the SDO/HMI longitudinal magnetogram acquired at 20:01:55 UTC (see Figure~\ref{fig:fov_obs}).  The subregion is selected so that the upward and downward flux near AR 11408 is approximately balanced.   We utilize the routine called `FFF' available in the SolarSoft IDL library to calculate the extrapolation based on the equations of \inlinecite{alissandrakis1981}  and \inlinecite{gary1989}.  By visual inspection alone, the observed intensity structure of the fibrils exhibit shear with respect to the potential magnetic field lines (see northwest area of superpenumbra in Figure~\ref{fig:field_extrap}).

Figure~\ref{fig:pot_field_shear} compares the extrapolated and inferred magnetic field vector for each point in our inverted map, at the assumed height of 1750 km.  Despite the presense of noise in the inverted quantities, the correlation of the field topology is favorable at all field strengths, with clearly more scatter at lower strengths (blue data points).  The inverted magnetic field strengths (assuming a unity filling factor) are generally stronger than the potential field extrapolated from the line-of-sight SDO/HMI measurements.  Deviations from the unity line in the azimuth comparison, especially at 50$^{\circ}$ and 320$^{\circ}$, indicate the presence of small amounts of azimuthal shear. However, a few systematic groups of solutions can be identified in the inverted results, which are a product of the high level of noise and the use of the DIRECT optimization algorithm.  Still, the majority of the weak field ({\it i.e.}, superpenumbral) inclinations cluster at highly inclined horizontal magnetic fields consistent with the potential field extrapolations.  However, the degree of scatter is too large to quantify the degree of magnetic non-potentiality.

\section{Discussion and Summary}

We have demonstated the use of the He {\sc i} triplet at 1083 nm to map and investigate the full vector magnetic field structure of an active region including both a sunspot and its superpenumbral canopy, wherein atomic-level polarization signatures dominate \citep{schad2013}.  Although the results are promising, one particular challenge faced by our inversions has been the probable influence of a non-symmetric radiation field within and near the outer edges of the sunspot. We neglect this symmetry breaking here, but we argue that the use of a Milne-Eddington model inside the sunspot infers the field parameters with $\approx 20\%$ accuracy.  Yet, with continued work, the symmetry breaking could be quite useful, as we suspect the $180^{\circ}$ ambiguity of the transverse Zeeman could be overcome by extending the statistical equilibrium equations used by {\sc P-Hazel} to include the nonzero K components of the radiation field tensor. This is due to the added constrain placed on the polarization formation by the Hanle effect and the anisotropies of pumping radiation.  The greatest challenge to accomplishing this, however, may be the proper determination of the radiation field tensor at the transition frequencies between each major term of the orthohelium atomic model. 

Our results favor a fairly homogenous view of the lateral structure of the magnetic field hosting the superpenumbral magnetic canopy, as in \inlinecite{giovanelli1982}.  Evidence is provided by both the spatial continuity in the observed polarization structure and the results of the {\sc Hazel} inversions.  While the relatively low number of fibrils/filaments observed in low-resolution observations of the super-penumbra can give the impression that the superpenumbra is quite inhomogenous, our results seem intuitive considering the great density of thin fibrils viewed in superpenumbral observations using high-resolution spectral-imagers like IBIS (see Paper \rom{1}).  As \inlinecite{judge2006} points out, the thermal fine-structure of the chromosphere can lead to a more complex notion of the chromosphere than the magnetic field structure supports.  In the low-$\beta$ environment of the upper chromosphere, the average force balance is controlled by the magnetic field, not the thermal structure.  Note, however, that while magnetic canopies may exhibit a high degree of homogenity locally, the orientation of the fibrils surrounding magnetic concentrations is influenced by the distribution of the magnetic field in the photosphere \citep{bala2004}. 

How fibrils become individuated by density and/or thermal perturbations remains an open question. \inlinecite{leenaarts2012} gave evidence using advanced 3D non-LTE radiative transfer calculations of H$\alpha$ that H$\alpha$ fibrils may be introduced by density enhancements aligned with the magnetic field in 3D MHD simulations of the chromosphere.  But fibril formation is difficult to constrain observationally, in part due to the difficulty of establishing fibril connectivity with other regions of the atmosphere.  Moreover, understanding dynamical activity of fibrils requires connectivity to be established.  Thermal stratifications derived using LTE-inversions of the Ca \rom{2} line at 854.2 nm by \inlinecite{beck2014} offer one tentative means to estabish connectivity with the photosphere, but magnetic inferences like those presented here are still hindered by observational noise at the outer endpoints of the fibril structures.  

On active region scales, we have demonstrated that maps of the chromospheric vector magnetic field that include superpenumbral canopy fields can be achieved by inverting the He {\sc i} triplet.  One goal of this study is to provide measurements of the magnetic field at the base of the corona, where the field is more force-free than in the photosphere. This may ease the challenges of extrapolating the coronal magnetic field so that various methods might fall into greater consistency than found by \inlinecite{derosa2009}.  However, the observational challenges are still significant, as the level of noise in these already advanced measurements is hindering.

In the near future, novel measurement methods ({\it e.g.}, \inlinecite{schad2014}) for measuring the He {\sc i} triplet polarization will help better reap its unique advantages.  Higher sensitivity measurements are necessary to exploit the full potential of these He {\sc i} diagnostics, including its height diagnostic capability. The level of noise in our current He {\sc i} observations may mask weak fine structuring of the magnetic field. The height-sensitivity of the He {\sc i} triplet at 1083 nm may be especially useful in light of the prediction by \inlinecite{leenaarts2012} that fibrils seen in the H$\alpha$ spectral line may on average have a greater formation height than the inter-fibril material. 


\appendix

\section{The Radiation Field Tensor Near a Sunspot}\label{append:field_tensor}

\begin{figure}
\centerline{\includegraphics[width=0.99\textwidth]{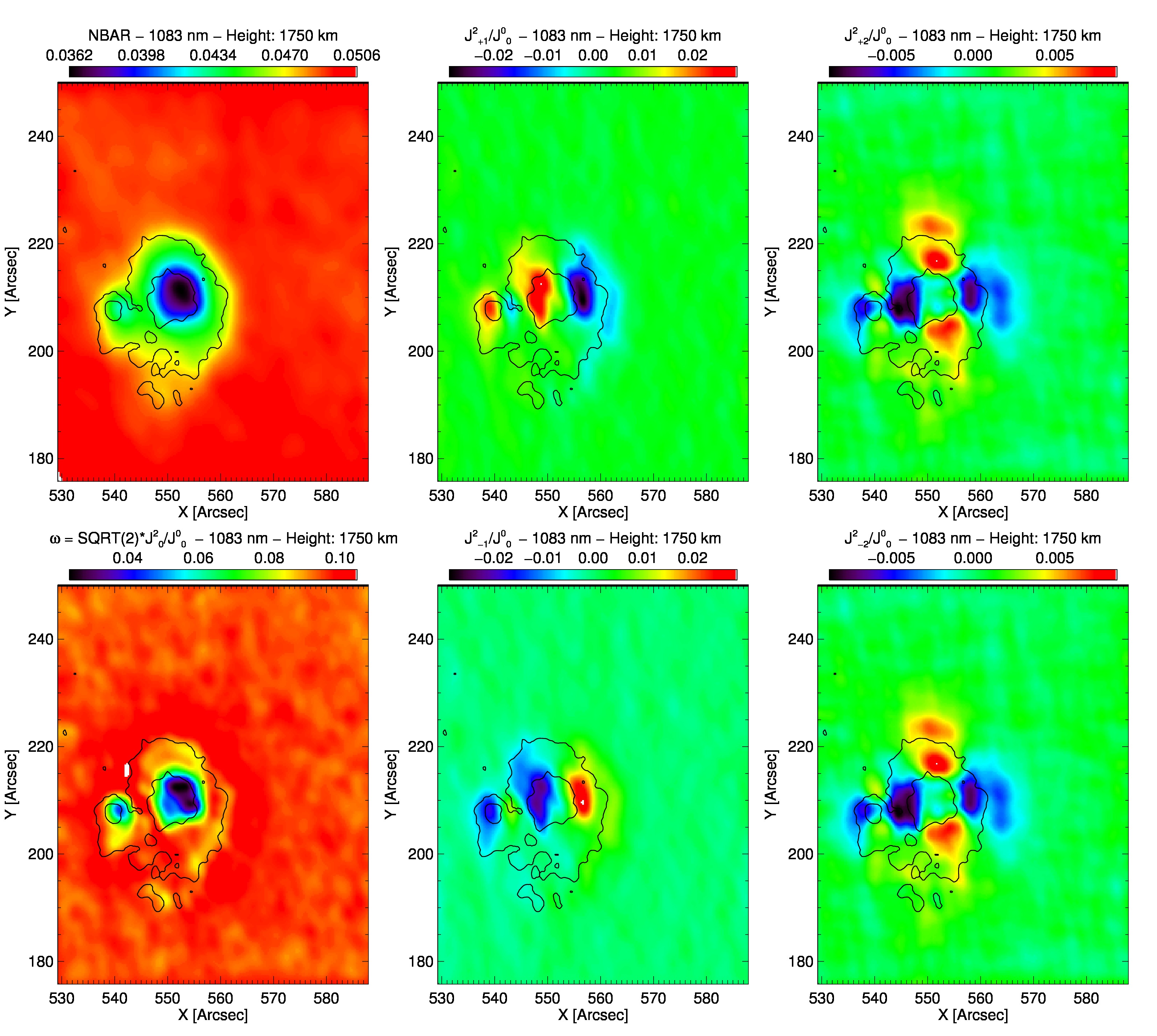}}
\caption{Multipole moments of the local radition field tensor for a slab of material located at a height of 1750 km above the solar surface.  Only the real part of $Q= \pm 1,2$ components are displayed.  NBAR represents the zero-moment, where NBAR is defined as $\bar{n} = (c^{2}/2h\nu^{3})J_{0}^{0}$.} 
\label{fig:rad_tensor}
\end{figure}

The pumping radiation field responsible for the development of population imbalances and quantum coherences in the orthohelium atomic system can be fully specified by the irreducible components of the statistical spherical tensor given in Equation~(\ref{eqn:rad_field_tensor}).  In the presence of symmetry breaking structure, the radiation field tensor can be determined by numerically integrating the radiation field given by the observed continuum structure of the region.  Our observations give the normalized continuum intensity near 1083 nm.  Thus, the absolute photometric intensity across the region can be inferred from the observations using the known limb-darkening law at 1083 nm.  Then, by transforming first to the local frame of reference of a given point in the atmosphere at height, h, we find the nonzero radiation field tensor components by applying following expressions assuming the radiation field is unpolarized, 
\begin{eqnarray}
J_{0}^{0}(\nu)        & = &                        \oint \frac{d\Omega}{4\pi}   		                   I(\nu,\vec{\Omega}) \\
J_{0}^{2}(\nu)        & = & \frac{1}{2\sqrt{2}}    \oint \frac{d\Omega}{4\pi} (3\cos^{2}\theta - 1)                I(\nu,\vec{\Omega})  \\
J_{\pm1}^{2}(\nu)     & = & \mp \frac{\sqrt{3}}{2} \oint \frac{d\Omega}{4\pi} \sin\theta \cos\theta e^{\pm i\chi}  I(\nu,\vec{\Omega})  \\
J_{\pm2}^{2}(\nu)     & = & \frac{\sqrt{3}}{4}     \oint \frac{d\Omega}{4\pi} \sin^{2}\theta e^{\pm 2 i\chi}       I(\nu,\vec{\Omega}),
\end{eqnarray}
which are the expansion of Equation~(\ref{eqn:rad_field_tensor}) for $i=0$. We follow the geometry used in Section 12.3 of \inlinecite{landi_2004}.  Iterated Gaussian quadrature bivariate integration is used to numerically integrate the above equations.  The results are illustrated in Figure~\ref{fig:rad_tensor} for a height of $h = 1750$ km.  Only the real portion of the $Q= \pm 1,2$ components are shown.

The symmetry breaking is most pronounced within the sunspot.  Contours of the inner and outer penumbral edges show the sunspot's photospheric extent (Figure~\ref{fig:rad_tensor}).  The mean intensity structure given by $\bar{n}$ (see Figure~\ref{fig:rad_tensor}) is slightly offset from the photospheric structure due to the observational geometry of the region, and the assumed height of the helium slab.  The weak ring structure in the map of $\omega$ best illustrates the extent of the radiative symmetry breaking in the superpenumbra.  The half angle of the heliographic extent of the light cone at this height is $4.06^{\circ}$, whereas the entire heliographic extent of the observed region is $4.44^{\circ} \times 5^{\cdot}$.  The impact of the $Q=\pm 1$ components is minimal outside of the sunspot's edge, whereas the $Q=\pm 2$ components show pronounced changes within the inner superpenumbral region.  However, the magnitude of these components are an order of magnitude below the $J^{2}_{0}$ component.  Figure~\ref{fig:rad_tensor}, of course, only shows the radiation field tensor at the frequency of one transition within the orthohelium atomic system.  Due to the variation of the Planck function, we might expect the magnitude of the $Q\ne 0$ components to increase at shorter wavelengths where the sunspot contrast is enhanced; yet, overall the impact of this symmetry breaking around the penumbral edge is expected to be an order of magnitude below the $J^{2}_{0}$ component which itself induces very weak polarization signatures of $\lesssim 0.1\%$ of the incoming intensity.

\begin{acks}

The NSO is operated by the Association of Universities for Research in Astronomy, Inc. (AURA), for the National Science Foundation.  FIRS has been developed by the Institute for Astronomy at the University of Hawai`i jointly with the NSO.  The FIRS project was funded by the National Science Foundation Major Research Instrument program, grant number ATM-0421582. We acknowledge the courtesy of the NASA/SDO HMI science team for providing high quality data.  We also extend our gratitude to Andres Asensio Ramos and Andreas Lagg for developing very useful inversion tools for the He {\sc i} triplet. 

\end{acks}


\bibliographystyle{spr-mp-sola-limited}
\bibliography{schad_bibliography}

\end{article} 
\end{document}